\pdfoutput=1

\newif\ifdebug\debugtrue
\debugfalse
\newif\ifreview\reviewtrue
\reviewfalse

\documentclass[
  sigconf,
  authorversion,
  natbib=false,
  anonymous=false,
  review=\ifreview true\else false\fi,
  authordraft=\ifdebug true\else false\fi,
  timestamp=\ifdebug true\else false\fi,
]{acmart}
\usepackage[utf8]{inputenc}

\RequirePackage[
  abbreviate=true,
  dateabbrev=true,
  isbn=true,
  doi=true,
  urldate=comp,
  url=true,
  maxbibnames=4,
  backref=false,
  backend=biber,
  style=ACM-Reference-Format,
  language=american
]{biblatex}
\usepackage{amsmath,amssymb}
\usepackage{booktabs}  
\usepackage{textcomp}  

\def\unif{\includegraphics[scale=1.1]{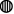}}
\def\Unif{\includegraphics[scale=1.4]{figs/U}}
\newcommand{\smath}[1]{\raisebox{.15ex}{\scalebox{.8}{#1}}}
\let\system\textrm 
\newcommand{\furl}[1]{\footnote{\url{#1}}}
\newcommand{\fref}[2]{#2\furl{#1}}

\setcopyright{acmlicensed}

\title{\textit{MIaS}: Math-Aware Retrieval in Digital Mathematical Libraries}

\author{Petr Sojka}
\orcid{0000-0002-5768-4007}
\affiliation{%
  \institution{Masaryk University}
  \department{Faculty of Informatics}
  \streetaddress{Botanick\'a 68a}
  \city{Brno}
  \state{Czech Republic}
  \postcode{602 00}
}
\email{sojka@fi.muni.cz}

\author{Michal R\r u\v zi\v cka}
\orcid{0000-0001-5547-8720}
\affiliation{%
  \institution{Masaryk University}
  \department{Faculty of Informatics}
  \streetaddress{Botanick\'a 68a}
  \city{Brno}
  \state{Czech Republic}
  \postcode{602 00}
}
\email{mruzicka@mail.muni.cz}

\author{V\'it Novotn\'y}
\orcid{0000-0002-3303-4130}
\affiliation{%
  \institution{Masaryk University}
  \department{Faculty of Informatics}
  \streetaddress{Botanick\'a 68a}
  \city{Brno}
  \state{Czech Republic}
  \postcode{602 00}
}
\email{witiko@mail.muni.cz}

\acmConference[CIKM '18]{%
  The 27th ACM International Conference on Information and Knowledge
  Management}{October 22--26, 2018}{Torino, Italy}
\acmBooktitle{%
  The 27th ACM International Conference on Information and Knowledge Management
  (CIKM '18), October 22--26, 2018, Torino, Italy}
\acmDOI{10.1145/3269206.3269233}
\acmISBN{978-1-4503-6014-2/18/10}
\acmPrice{15.00}
\acmYear{2018}
\copyrightyear{2018}

\keywords{%
  Math Information Retrieval,
  Digital Mathematical Libraries
}

\ifdebug
\usepackage[normalem]{ulem}
\usepackage[hybrid,citations]{markdown}

\fi
\begin{document}

\begin{abstract}
Digital mathematical libraries (DMLs) such as arXiv, Numdam,
and \href{https://eudml.org}{EuDML} contain mainly documents from STEM fields,
where mathematical formulae are often more important than text for understanding.
Conventional information retrieval (IR) systems are unable to
represent formulae and they are therefore ill-suited for math information
retrieval (MIR).
%
To fill the gap, we have developed, and open-sourced the \system{MIaS} MIR system.
%
MIaS is based on the full-text search engine Apache Lucene. On top of text
retrieval, MIaS also incorporates a set of tools for preprocessing
mathematical formulae.
%
We describe the design of the system and present speed, and quality evaluation
results.
%
We show that MIaS is both efficient, and effective, as evidenced by our
victory in the NTCIR-11 Math-2 task.\looseness=-1
\end{abstract}

\maketitle

\section{Introduction}
In mathematical discourse, formulae are often more important than text for
understanding. As a result, digital mathematical libraries (DMLs) require
math information retrieval (MIR) systems that recognize both text and math in
documents and queries.
Conventional IR systems represent both text, and formulae using the
bag-of-words vector-space model (VSM). However, the VSM captures neither the
structural, nor the semantic similarity between mathematical formulae, which
makes it ill-suited for MIR.

To fill the gap, new math-aware IR systems started to appear after the
pioneering workshop on DMLs~\cite{dml:dml2008proceedings}. Springer's
\system{\LaTeX{}
Search}\furl{https://www.ams.org/notices/201004/rnoti-apr10-cov4.pdf} system
takes formulae from papers with available \LaTeX{} sources, and hashes the formulae to obtain a text representation.
Zentral\-blatt Math uses the \system{MathWebSearch}
system\furl{https://zbmath.org/formulae/}~\cite{kohlhase2008mathwebsearch},
which represents formulae with substitution trees. We have
developed and open-sourced the MIaS (Math Indexer and Searcher)
system\furl{https://github.com/MIR-MU/MIaS}~\cite{mir:MIaSNTCIR-11,Ruzicka17Math}
using the robust highly-scalable full-text search engine 
\href{https://lucene.apache.org/}{Apache Lucene}~\cite{bialecki12} 
and our own set of tools for the preprocessing
of mathematical formulae. Since 2012, MIaS has been deployed in 
\fref{https://eudml.org/search}{the European
Digital Mathematical Library (EuDML)},
making it historically the first system to be deployed in a~DML.


\begin{figure*}[tbh]
\centering
\includegraphics[width=\textwidth]{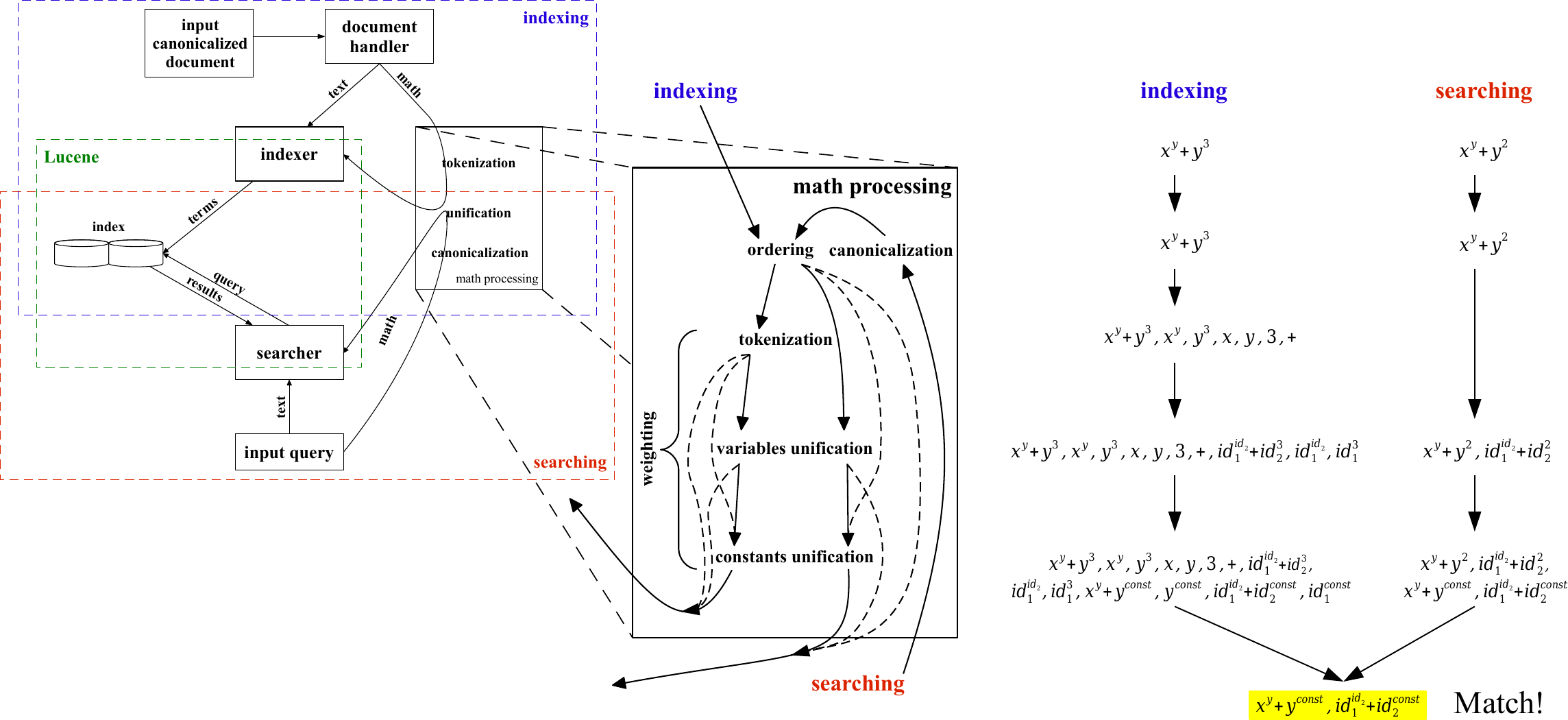}
\caption{The preprocessing of mathematical formulae in indexed and query documents}
\label{fig:system}
\end{figure*}

\section{System Description}
MIaS processes text and math separately.
The text is tokenized and stemmed to unify inflected word forms.
Math is expected to be in \fref{https://www.w3.org/TR/MathML3/}{the MathML format}
. 
Open tools such as \fref{https://www-sop.inria.fr/marelle/tralics/}{Tralics}
, 
\fref{https://dlmf.nist.gov/LaTeXML/}{\LaTeX ML}
convert documents in the popular math
authoring language of \LaTeX{} to MathML.
Other tools such as \href{http://www.inftyreader.org/}{InftyReader}~\cite{dml:suzukietal2003}
, and
\href{https://github.com/zorkow/MaxTract}{MaxTract}~\cite{DBLP:conf/aisc/BakerSS12} 
convert raster, and vector PDF documents, respectively, to MathML.
The math is then canonicalized, ordered, tokenized, and unified (see Figure~\ref{fig:system}). We will describe each of these processing steps in detail in the following paragraphs.

\paragraph{Canonicalization}
As explained above, MathML can originate from multiple sources and each can
encode equivalent mathematical formulae a little differently. To obtain a
single \emph{canonical} representation, we initially used the third-party
MathML canonicalizer from the UMCL library that converts math to a subset of
MathML called the Canonical MathML~\cite{acc:archambaultmoco06cmathml}.
However, since the conversion speed and accuracy did not match our
expectations, we have developed and open-sourced our own MathML
canonicalizer\footnote{\url{https://github.com/MIR-MU/MathMLCan}}~%
\cite{FormanekEtAl:OpenMathUIWiP2012}.

\paragraph{Ordering}
MathML canonicalization only affects the encoding of mathematical formulae and
does not result in any syntactic manipulation. We go a step further and reorder
the operands of commutative operators alphabetically. For example, we convert
the formulae $a+b$, and $b+a$ to a single canonical form $a+b$.\looseness=-1

\paragraph{Tokenization}
A user of our system may not know the precise form of a formula they are
searching for. To enable partial matches, we index not only the original formula,
but also all its \emph{subformulae}, which correspond to all the XML subtrees
of the original formula XML tree. To penalize partial matches, the weight of
subformulae is inversely proportional to their depth in the XML
tree.~\cite{dml:sojkaliska2011}

A user is likely interested in documents that contain either the query
formula itself, or larger formulae with the query formula as a subformula.
On the other hand, a user is unlikely to be interested in documents that
contain only small parts of the query formula, such as isolated numbers, and
symbols. For that reason, we only tokenize formulae in indexed documents, not
in user queries.

\paragraph{Unification}
In theory, the naming of variables does not affect the meaning of formulae. To
match formulae in different notations, we replace each variable with a
numbered identifier. For example, we convert the formulae $a+b^a,$ and $x+y^x$
to a single \emph{unified} form $\text{id}_1+\text{id}_2^{\text{id}_1}$. In
practice, many fields have an established notation and variable names are
meaningful. To encourage precise matches, we keep the original formulae in
addition to the unified formulae.

Two formulae that only differ in numeric constants are often related. For
example, both $3x^2-2x+2,$ and $8x^2-3x+6$ are quadratic
polynomials. We replace every numeric constant with a constant identifier. For
example, we convert the above formulae to a single unified form
$\text{const}x^2-\text{const}x+\text{const}$. To encourage precise matches, we
keep the original formulae in addition to the unified formulae.

In predicate logic, a variable can represent an arbitrary formula. For example,
the formulae $a^2+{}$\smath{$\frac{\sqrt b}{c}$}, and $a^2+{}$\smath{$\frac
xy$} are equivalent if $x$ equals $\sqrt b$. Starting with the deepest
subformulae, we replace all subformulae at a given depth with a unifying
identifier.~\cite{RuzickaSojkaLiska16Math} For example, we convert the formula
$a^2+{}$\smath{$\frac{\sqrt
b}{c}$} to a sequence of \emph{structurally unified} formulae
$a^2+{}$\smath{$\frac{\sqrt{\unif}}{c}$}$,\Unif^{\unif}+{}$\smath{$\frac\unif\unif$}, and
$\Unif+\Unif$ and the formula $a^2+{}$\smath{$\frac x y$} to a sequence of
structurally unified formulae $\Unif^{\unif}+{}$\smath{$\frac\unif\unif$}, and
$\Unif+\Unif$. To penalize partial matches, the weight of the formulae is
proportional to the depth of replacement. To encourage precise matches, we keep
the original formulae in addition to the unified formulae. We have open-sourced
\fref{https://github.com/MIR-MU/MathMLUnificator}{the MathML structural unificator}.

\begin{figure}
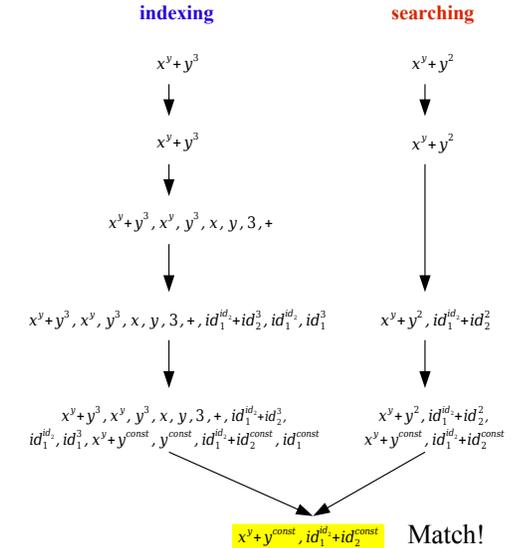

\begin{tabular}{lccccc}
Subquery 1: & $f_1$ & $f_2$ & $t_1$ & $t_2$ & $t_3$ \\
Subquery 2: & $f_1$ & $f_2$ & $t_1$ & $t_2$ &       \\
Subquery 3: & $f_1$ & $f_2$ & $t_1$ &       &       \\
Subquery 4: & $f_1$ & $f_2$ &       &       &       \\
Subquery 5: & $f_1$ &       & $t_1$ & $t_2$ & $t_3$ \\
Subquery 6: &       &       & $t_1$ & $t_2$ & $t_3$ \\
\end{tabular}
\caption{The subqueries produced from the original query $f_1 f_2 t_1 t_2 t_3$
  with mathematical formulae $f_1$, and $f_2$ and terms $t_1, t_2,$ and
  $t_3$ using the Leave Rightmost Out (LRO) strategy.}
\label{fig:query-expansion}
\end{figure}

\makeatletter\let\@adddotafter\relax\paragraph{}\makeatother
After preprocessing, a query consists of a weighted set of terms, and formulae.
Since we are now going to search for documents that match at least one term,
and at least one formula from the query, ill-posed terms, and formulae will
negatively impact the recall of our system. To overcome this problem, we remove
selected terms and formulae to produce a set of \emph{subqueries}.
Figure~\ref{fig:query-expansion} shows an example strategy for producing
subqueries. \textcite{LiskaSojkaRuzicka15Combining} describe other strategies
that we use. We then submit the subqueries to Apache Lucene and receive ranked
lists of resulting documents. Since the scores of the resulting documents are
incomparable between subqueries, we cannot merge and rerank the individual
result lists. Instead, we interleave them to obtain the final search results
that we present to the user.

\begin{figure*}
\includegraphics[width=\textwidth]{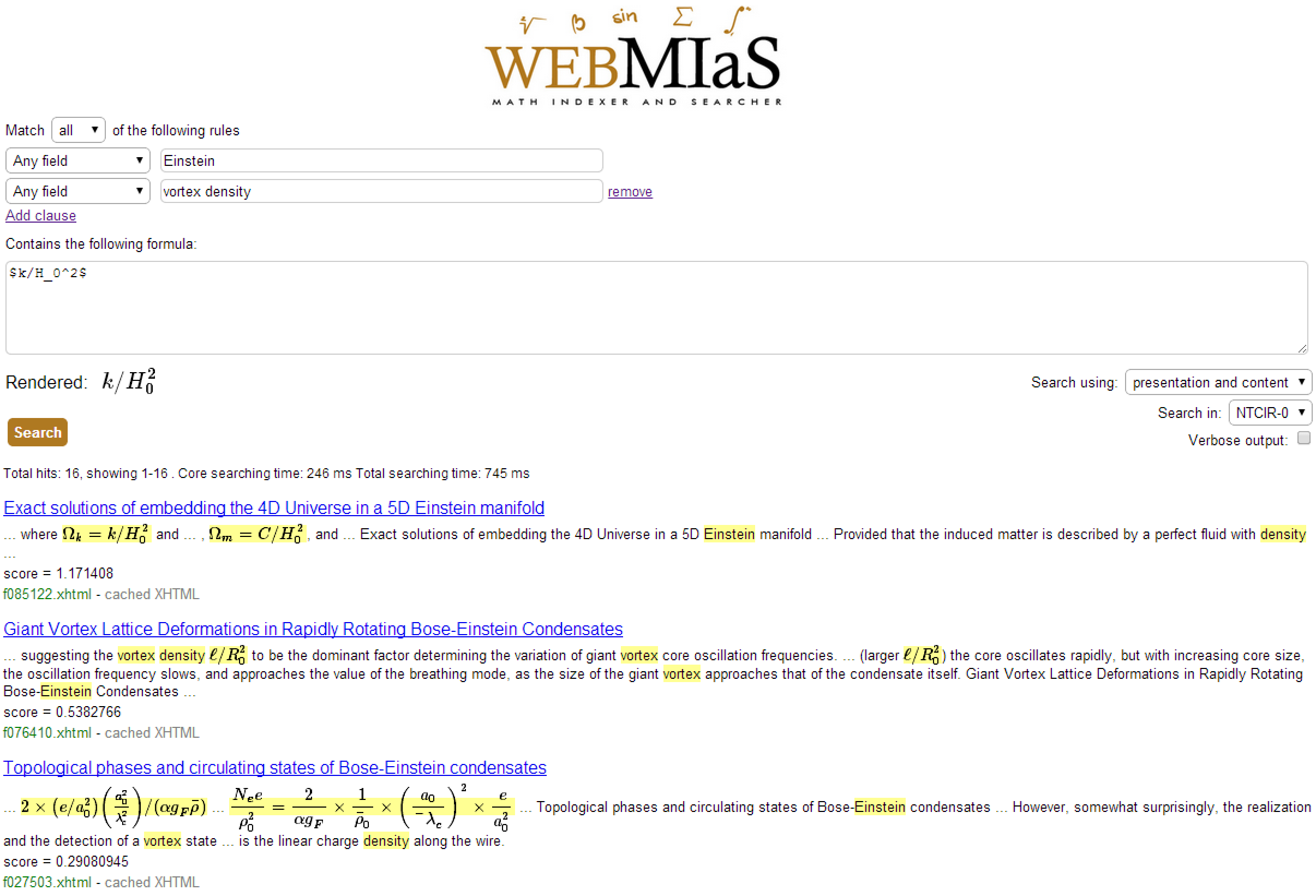}
\caption{The user interface of WebMIaS. Users can input their query in a
  combination of text, and math with native support for \LaTeX{} provided by
  Tralics, and MathJax. Matches are conveniently highlighted in the search
  results.}
\label{fig:webmias}
\end{figure*}

\begin{table}
\caption{Speed evaluation results on the MREC dataset using 448G of RAM, and
  eight Intel Xeon\texttrademark{} X7560 2.26\,GHz CPUs.\looseness=-1}
\label{tab:speed-eval-mrec}
\begin{tabular}{rrrrr}
& \multicolumn{2}{r}{Mathematical (sub)formulae} & \multicolumn{2}{r}{Indexing time (min)} \\
Docs & Input & Indexed & Real & CPU \\ \midrule
10,000 &
 3,406,068 &
 64,008,762 &
 35.75 &
 35.05 \\
50,000 &
 18,037,842 &
 333,716,261 &
 189.71 &
 181.19 \\
100,000 &
 36,328,126 &
 670,335,243 &
 384.44 &
 366.54 \\
200,000 &
 72,030,095 &
 1,326,514,082 &
 769.06 &
 733.44 \\
300,000 &
 108,786,856 &
 2,005,488,153 &
 1,197.75 &
 1,116.64 \\
350,000 &
 125,974,221 &
 2,318,482,748 &
 1,386.66 &
 1,298.10 \\
439,423 &
 158,106,118 &
 2,910,314,146 &
 1,747.16 &
 1,623.22 \\
\end{tabular}

\vspace{2.4em}
  \caption{Speed evaluation results on the NTCIR-11 Math-2 dataset using
  the same computer as above.}
\label{tab:speed-eval-ntcir11}
\begin{tabular}{rrrrr}
& \multicolumn{2}{r}{Mathematical (sub)formulae} & \multicolumn{2}{r}{Indexing time (min)} \\
Docs & Input & Indexed & Real & CPU \\ \midrule
8,301,545 & 59,647,566 & 3,021,865,236 & 1940.07 & 3,413.55
\end{tabular}
\end{table}

\makeatletter\let\@adddotafter\relax\paragraph{}\makeatother
To provide a web user interface to MIaS, we have developed and
open-sourced \fref{https://mir.fi.muni.cz/webmias/}{WebMIaS}$^,$%
\furl{https://github.com/MIR-MU/WebMIaS}~\cite{mir:MIaSNTCIR-11,
dml:cicm2014liskaetal}. Users can input their query in a
combination of text, and math with a native support for \LaTeX{} provided by
Tralics, and \href{https://www.mathjax.org/}{MathJax}~\cite{cervone2012mathjax}.
Matches are conveniently highlighted in the search results.
The user interface of WebMIaS is shown in Figure~\ref{fig:webmias}.
We have deployed a demo of \fref{https://mir.fi.muni.cz/webmias-demo/}{the
latest development version of WebMIaS} using the 
\fref{https://tomcat.apache.org/}{Apache
Tomcat} implementation of the Java
Servlet. The demo uses an index built from a subset of the arXMLiv
dataset~\cite{dml:arXMLiv2010} made available to the NTCIR-12 conference
participants and will serve as the basis for our live demonstration at the
conference.

\section{Evaluation}
We performed a speed evaluation of MIaS on the MREC dataset of 439,423
documents~\cite{dml:liska2011} (see Table~\ref{tab:speed-eval-mrec}), a quality
and speed evaluation on the NTCIR-10~Math~\cite{mir:NTCIR-10-Overview,
MIR:MIRMU} dataset of 100,000 documents, and a quality and speed evaluation on
the NTCIR-11~\mbox{Math-2}~\cite{NTCIR11Math2overview,mir:MIaSNTCIR-11} (see
Tables~\ref{tab:speed-eval-ntcir11}, and \ref{tab:quality-eval-ntcir11}), and
NTCIR-12~MathIR~\cite{ZanibbiEtAl16NTCIR,RuzickaSojkaLiska16Math} dataset of
105,120 documents that were split into 8,301,578 paragraphs. Speed evaluation
shows that the indexing time of our system is linear in the number of indexed
documents and that the average query time is 469\,ms.  With respect to quality
evaluation, MIaS has notably won the NTCIR-11~Math-2 task.

\begin{table}
\caption{Quality evaluation results on the NTCIR-11 Math-2 dataset. The mean
  average precision (MAP), and precisions at ten (P@10), and five (P@5) are
  reported for queries formulated using Presentation (PMath), and Content
  MathML (CMath), a combination of both (PCMath), and \LaTeX. Two different
  relevance judgement levels of $\geq1$ (partially relevant), and $\geq3$
  (relevant) were used to compute the measures. Number between slashes (/$\cdot$/)
  is our rank among all teams.}
\label{tab:quality-eval-ntcir11}
\vspace*{-.5ex}
\begin{tabular}{llllll}
Measure & Level & PMath & CMath & PCMath & \LaTeX \\ \midrule
  MAP & 3 & 0.3073 & \textbf{0.3630 /1/} & 0.3594 & 0.3357 \\
  P@10 & 3 & 0.3040 & \textbf{0.3520 /1/} & 0.3480 & 0.3380 \\
  P@5 & 3 & 0.5120 & \textbf{0.5680 /1/} & 0.5560 & 0.5400 \\
  MAP & 1 & 0.2557 & \textbf{0.2807 /2/} & 0.2799 & 0.2747 \\
  P@10 & 1 & 0.5020 & 0.5440 & \textbf{0.5520 /1/} & 0.5400 \\
  P@5 & 1 & 0.8440 & \textbf{0.8720 /2/} & 0.8640 & 0.8480 \\
\end{tabular}
\end{table}

\section{Conclusion and Future Work}
With the growing importance of DMLs, there is a growing demand for effective MIR systems.
%
The evaluation shows that our open-source \system{MIaS} system is both efficient, and effective while building on industrial-strength full-text search engine Apache Lucene.
The system allows low-latency responses even on the big math corpora as proved
by its deployment in EuDML.

The speed of indexing and response latency of MIR will be further increased by the migration of \system{MIaS} from Apache Lucene to the distributed full-text search engine \fref{https://elastic.co}{ElasticSearch}.
%
The idea of indexing structures rather than terms can be generalized from mathematical formulae to semi-structured text.
Reordering the operands of associative operators is only a simple transformation.
For example, to convert $\sqrt[n]{a}$, and $a^{1/n}$ to a single canonical representation, a general computer algebra system (CAS) can be used.
We experiment~\cite{rygletal16} with improving the vector space representations of document passages, aiming to add support for mathematics in the future.
Embeddings can also be computed for equations~\cite{mir:krstowski2018arXiv} now, which presents new possibilities of using language modeling for the semantic segmentation of STEM articles, and weighting the segments~\cite{rygletal16}.
Grasping the meaning of mathematical formulae is crucial: content is king.
\looseness=-1

\subsubsection*{Acknowledgements}
We gratefully acknowledge the support by the European Union under the FP7-CIP program,
project 250,503 (EuDML), and by the ASCR under the Information Society R\&D program, project 1ET200190513 (DML-CZ).
We also sincerely thank three anonymous reviewers for their insightful comments.


\printbibliography

\end{document}